\documentclass[12pt]{article}
\newcommand{\be}{\begin{equation}}
\newcommand{\ee}{\end{equation}}
\newcommand{\bea}{\begin{eqnarray}}
\newcommand{\eea}{\end{eqnarray}}
\newcommand{\sptwo}{1.4}
\newcommand{\doublespace}{\edef\baselinestretch{\sptwo}\Large\normalsize}
\newcommand{\newsection}[1]{\section{#1}\setcounter{equation}{0}}

\newcounter{newapp}
\begin{document}
\vspace*{0.2in}
\begin{center}
{\large\bf Gauging Nonlinear Supersymmetry}\\
\end{center}
\vspace{0.2in}
\begin{center}
{T.E. Clark}\footnote{e-mail address: clark@physics.purdue.edu}$~^a~~,~~${S.T. Love}\footnote{e-mail address: loves@physics.purdue.edu}$~^{a}~~,~~${Muneto Nitta}\footnote{e-mail address: nitta@th.phys.titech.ac.jp}$~^b~~,~~${T. ter Veldhuis}\footnote{e-mail address: terveldhuis@macalester.edu}$~^c$\\
\end{center}
~\\
~\\
a. {\it Department of Physics, Purdue University, West Lafayette, IN 47907-1396}\\
~\\
b. {\it Department of Physics, Tokyo Institute of Technology, Tokyo 152-8551, Japan}\\
~\\
c. {\it Department of Physics \& Astronomy, Macalester College, Saint Paul, MN 55105-1899}
\vspace{0.2in}
~\\
~\\
~\\
\begin{center}
{\bf Abstract}
\end{center}
Coset methods are used to construct the action describing the dynamics associated with the spontaneous breaking of the local supersymmetries.  The resulting action is an  invariant form of the Einstein-Hilbert action, which in addition to the gravitational vierbein, also includes a massive gravitino field.  Invariant interactions with matter and gauge fields are also constructed.  The effective Lagrangian describing processes involving the emission or absorption of a single light gravitino is analyzed.
~\\
~\\
~\\
\pagebreak
\doublespace

\newsection{Introduction}

The low energy dynamics of a theory with spontaneously broken global supersymmetry  includes the Nambu-Goldstone fermion, the Goldstino, of the broken SUSY.  The action for the Goldstino fields, denoted by Weyl spinors $\lambda_\alpha$ and $\bar\lambda_{\dot\alpha}$, with $\alpha = 1,2$ and $\dot\alpha =1,2$, was found by Akulov and Volkov \cite{Volkov:jx} to be
\be
\Gamma = -F^2 \int d^4 x \det{A} = -F^2 \int d^4 x \det{(\delta_\mu^{~m} -i\lambda  \stackrel{\leftrightarrow}{\partial}_\mu \sigma^m \bar\lambda)},
\ee
where $A_\mu^{~m} = (\delta_\mu^{~m}-i\lambda  \stackrel{\leftrightarrow}{\partial}_\mu \sigma^m \bar\lambda )$ is the Akulov-Volkov vierbein  and $F^2$ is the Goldstino decay constant, a scale set by the dynamics responsible for SUSY breaking. When the supersymmetry is realized as a local symmetry, the super-Higgs mechanism becomes operational and the Goldstino provides the spin $\frac{1}{2}$ components of the now massive spin $\frac{3}{2}$ gravitino fields, which are denoted by $\psi_{\mu\alpha}$ and $\bar\psi_{\mu\dot\alpha}$ with $\mu =0, 1, 2, 3$.  On  the other hand, the space-time coordinate invariance remains unbroken and so the graviton, which is described by the vierbein $e_\mu^{~m}$, with $m= 0, 1, 2, 3$, remains a massless spin $2$ field \cite{Volkov:1973jd}-\cite{Baulieu:1985wa}.  The purpose of this paper is to derive the low energy action governing the dynamics of these degrees of freedom and then examine some of its consequences.  This is achieved using the method of nonlinear realizations \cite{Coleman:sm}.  In section 2, this coset method is applied to the case of the local super-Poincar\'e group ${\cal SP}_4$ \cite{Volkov:1973jd}, \cite{VolkovSoroka2}.  The generalized locally covariant Maurer-Cartan one-form is constructed. It includes the vierbein as well as  the locally covariant derivatives of the Goldstino fields which involve the gravitino fields and the spin (and hence affine) connections.  The general covariant derivatives of these covariant one-forms are used as the building blocks of the action \cite{Utiyama:1956sy}.  In section 3, the invariant action for spontaneously broken supergravity is constructed from the above mentioned Maurer-Cartan one-forms and their derivatives.  The action is transformed to the unitary gauge which more clearly reveals the physical content of the supergravity vierbein and massive gravitino.  Note that the nonlinear realization of local symmetry which we construct is achieved using only the graviton and gravitino degrees of freedom. There is no need to include other degrees of freedom which appear in linear realizations of supergravity.  In section 4, the couplings to matter and gauge fields are also determined using the covariant one-forms.  In particular, the interactions of the gravitino with the Standard Model particles describing its single emission or absorption are delineated.  These match those obtained previously \cite{Antoniadis:2004se} using the equivalence theorem \cite{Casalbuoni:1988qd} to describe high energy processes involving the helicity $\pm \frac{1}{2}$ components of the gravitino.

\newsection{The Coset Construction}

In this section, we construct the nonlinear realization of the super-Poincar\'e group of transformations ${\cal SP}_4$ when it is spontaneously broken to the Poincar\'e subgroup ${\cal P}_4$.  The method of nonlinear realizations begins with the construction of the coset element $\Omega \in {\cal SP}_4/SO(1,3)$
\be
\Omega (x) = e^{ix^\mu  P_\mu} e^{i[\lambda^\alpha (x) Q_\alpha + \bar\lambda_{\dot\alpha} (x) Q^{\dot\alpha} ]},
\ee
where the $SO(1,3)$ subgroup is the Lorentz stability group.  The coset elements are labelled by the space-time coordinates $x^\mu$ and the superspace coordinates 
$\lambda_\alpha (x)$ and $\bar\lambda_{\dot\alpha}(x)$, which are the Weyl spinor Goldstino fields.  The generators of ${\cal SP}_4$ are the energy-momentum operator $P_\mu$, the supersymmetry Weyl spinor charges $Q_\alpha$ and $\bar{Q}_{\dot\alpha}$ and the angular momentum operator $M_{\mu\nu}$.  They obey the usual SUSY algebra
\bea
\left[M_{\mu\nu} , M_{\rho\sigma} \right] &=& -i \left( \eta_{\mu\rho} M_{\nu\sigma} -\eta_{\mu\sigma} M_{\nu\rho} +\eta_{\nu\sigma} M_{\mu\rho} -\eta_{\nu\rho} M_{\mu\sigma} \right) \cr
\left[M_{\mu\nu} , P_{\lambda} \right] &=& i \left( P_\mu \eta_{\nu\lambda} - P_\nu \eta_{\mu\lambda} \right) \cr
\left[M_{\mu\nu} , Q_\alpha \right] &=& -\frac{1}{2}\left(\sigma^{\mu\nu} \right)_\alpha^{~\beta} Q_\beta \cr
\left[M_{\mu\nu} , \bar{Q}_{\dot\alpha} \right] &=& \frac{1}{2}\left(\bar\sigma^{\mu\nu} \right)_{\dot\alpha}^{~\dot\beta} \bar{Q}_{\dot\beta} \cr
\left\{Q_\alpha , \bar{Q}_{\dot\alpha} \right\} &=& {2}\sigma^{\mu}_{\alpha\dot\alpha} P_\mu ,
\label{Algebra}
\eea
with all remaining commutators vanishing. Here $\eta_{\mu\nu}$ is the Minkowski space metric tensor defined with signature $(+1, -1, -1, -1)$.

Left multiplication of $\Omega$ by an ${\cal SP}_4$ group element $g$ characterized by the local infinitesimal parameters $\epsilon^\mu (x), \xi_\alpha (x), \bar\xi_{\dot\alpha} (x), \alpha^{\mu\nu}(x)$ so that 
\be
g(x) = e^{i\epsilon^\mu(x) P_\mu} e^{i\xi^\alpha (x) Q_\alpha} e^{i\bar\xi_{\dot\alpha} (x) \bar{Q}^{\dot\alpha}} e^{\frac{i}{2} \alpha^{\mu\nu}(x) M_{\mu\nu}},
\ee
results in transformations of the space-time coordinates and the Nambu-Goldstone fields according to the general form \cite{Coleman:sm}
\be
g(x)\Omega(x) = \Omega^\prime(x^\prime) h(x) .
\label{leftmult}
\ee
The transformed coset element, $\Omega^\prime$,  is a function of the transformed space-time coordinates and the total variations of the fields so that
\be
\Omega^\prime (x^\prime)= e^{ix^{\prime\mu}  P_\mu} e^{i[\lambda^{\prime \alpha} (x^\prime) Q_\alpha + \bar\lambda^{\prime}_{\dot\alpha} (x^\prime)\bar{Q}^{\dot\alpha}]}.
\ee 
In the ${\cal SP}_4$ case, $h$ is field independent and is simply an element of the Lorentz subgroup $SO(1,3)$ given by
\be
h= e^{\frac{i}{2} \alpha^{\mu\nu}(x) M_{\mu\nu}} .
\label{hele}
\ee
Exploiting the algebra of the ${\cal SP}_4$ charges displayed in equation (\ref{Algebra}), along with use of the Baker-Campbell-Hausdorff formulae, the local ${\cal SP}_4$ transformations are obtained as coordinate variations and total variations of the fields
\bea
x^{\prime\mu}  &=&  x^\mu +\Delta x^\mu 
 = x^\mu +\epsilon^\mu (x) +i[\xi (x) \sigma^\mu \bar\lambda (x) - \lambda (x) \sigma^\mu \bar\xi (x) ] -\alpha^{\mu\nu} (x) x_\nu \cr
\lambda_\alpha^\prime (x^\prime) &=& \lambda_\alpha (x) +\Delta \lambda_\alpha (x) =\lambda_\alpha (x) + \xi_\alpha (x) +\frac{i}{4} \alpha_{\mu\nu} (x)\left(\sigma^{\mu\nu} \right)_\alpha^{~~\beta} \lambda_\beta (x)\cr
\bar\lambda_{\dot\alpha}(x^\prime) &=& \bar\lambda_{\dot\alpha} (x) + \Delta \bar\lambda_{\dot\alpha} (x)= \bar\lambda_{\dot\alpha} (x) +\bar\xi_{\dot\alpha} (x)+\frac{i}{4} \alpha_{\mu\nu}\left(\bar\sigma^{\mu\nu} \right)_{\dot\alpha\dot\beta} \bar{\lambda}^{\dot\beta} (x) .
\label{variations}
\eea
The spontaneously broken SUSY transformations are nonlinearly realized as intrinsic variations of the fields, $\delta \lambda = \Delta \lambda -\Delta x^\mu \partial_\mu \lambda$ with analogous results for $\bar\lambda$.  The Nambu-Goldstone fields $\lambda_{\alpha}$ and $\bar\lambda_{\dot\alpha}$ transform inhomogeneously under the broken local SUSY transformations $Q_\alpha$ and $\bar{Q}_{\dot\alpha}$, respectively.  Thus these broken transformations can be used to transform to the unitary gauge in which both $\lambda_{\alpha}$ and $\bar\lambda_{\dot\alpha}$ vanish.  This will be done in section 3 in order to exhibit the physical degrees of freedom in a more transparent fashion.

The ${\cal SP}_4$ transformations induce a coordinate and field dependent general coordinate transformation of the space-time coordinates.  From the $x^\mu$ coordinate transformation given above, the general coordinate Einstein transformation for the space-time coordinate differentials is given by
\be
dx^{\prime \mu} = dx^\nu {G}_\nu^{~\mu} (x),
\label{dxprime}
\ee
with ${G}_\nu^{~\mu}(x) = \partial x^{\prime \mu}/\partial x^\nu$.  The ${\cal SP}_4$ invariant interval can be formed using the metric tensor ${g}_{\mu\nu}(x)$ so that $ds^2 = dx^\mu {g}_{\mu\nu}(x) dx^\nu = ds^{\prime 2} = dx^{\prime \mu} {g}^\prime_{\mu\nu}(x^\prime) dx^{\prime \nu}$ where the metric tensor transforms as 
\be
{g}^\prime_{\mu\nu} (x^\prime) = {G}_\mu^{-1\rho}(x) {g}_{\rho\sigma}(x) {G}_\nu^{-1\sigma}(x) .
\label{gprime}
\ee

The form of the vierbein (and hence the metric tensor) as well as the locally 
covariant derivatives of the Goldstino fields and the spin connection can be extracted from the locally covariant Maurer-Cartan one-form, $\omega \equiv \Omega^{-1}D\Omega$, which can be expanded in terms of the generators as 
\bea
\omega = \Omega^{-1} D\Omega &\equiv& \Omega^{-1} (d +i \hat{E})\Omega \cr
 &=& i\left[ \omega^m P_m + \omega_{Q}^\alpha Q_\alpha + \bar\omega_{\bar{Q}\dot\alpha} \bar{Q}^{\dot\alpha} +\frac{1}{2}\omega_M^{mn} M_{mn}\right].
\eea
Here Latin indices $m,n = 0,1, 2, 3$, are used to distinguish tangent space local Lorentz transformation properties from space-time Einstein transformation properties which are denoted using Greek indices. In what follows, Latin indices are raised and lowered using of the Minkowski metric tensors, $\eta^{mn}$ and $\eta_{mn}$, while Greek indices are raised and lowered with use of the curved metric tensors, $g^{\mu\nu}$ and $g_{\mu\nu}$.  Since the Nambu-Goldstone fields vanish in the unitary gauge it is useful to exhibit the one-form gravitational fields in terms of their translated form
\be
\hat{E} = e^{ix^\mu P_\mu} E e^{-ix^\mu P_\mu} .
\ee
The one-form gravitational fields $E$ have the expansion in terms of the charges as
\be
E= E^m P_m +\psi^\alpha Q_\alpha + \bar\psi_{\dot\alpha} \bar{Q}^{\dot\alpha} +\frac{1}{2}\gamma^{mn} M_{mn} .
\ee
Similarly expanding $\hat{E}$ as 
\be
\hat{E}= \hat{E}^m P_m + \hat\psi^\alpha Q_\alpha + \bar{\hat\psi}_{\dot\alpha} \bar{Q}^{\dot\alpha}+\frac{1}{2}\hat{\gamma}^{mn} M_{mn} ,
\ee
one finds the various fields are related according to 
\bea
\hat{E} &=& E^m +\gamma^{mn}x_n \cr
\hat\psi^\alpha &=& \psi^\alpha  \cr
\bar{\hat\psi}_{\dot\alpha} &=& \bar\psi_{\dot\alpha} \cr
\hat{\gamma}^{mn} &=& \gamma^{mn} . 
\eea

Defining the one-form gravitational fields to transform as a gauge field so that
\be
\hat{E}^\prime (x^\prime) = g(x)\hat{E}(x) g^{-1}(x) -ig(x)dg^{-1}(x),
\ee
the covariant Maurer-Cartan one-form transforms analogously to the way it varied for global transformations:
\be
\omega^\prime(x^\prime) = h(x)\omega (x) h^{-1}(x) +h(x)dh^{-1}(x),
\ee
with $h= e^{\frac{i}{2}\alpha^{mn}(x) M_{mn}}$.  Expanding in terms of the ${\cal SP}_4$ charges, the individual one-forms transform according to their local Lorentz nature as
\bea
\omega^{\prime m}(x^\prime) &=&  \omega^n (x)\Lambda^{~m}_{n}(\alpha (x)) \cr
\omega_{Q\alpha}^\prime (x^\prime) &=& D_{\alpha}^{(\frac{1}{2},0)\beta} (\alpha (x)) \omega_{Q\beta} \cr
\bar\omega_{\bar{Q}}^{\prime\dot\alpha} (x^\prime) &=& D_{~~~~~~~\dot\beta}^{(0,\frac{1}{2})\dot\alpha} (\alpha (x)) \bar\omega_{\bar{Q}}^{\dot\beta} \cr
\omega^{\prime mn}_M (x^\prime)&=& \omega^{rs}_M (x)\Lambda^{~m}_{r}(\alpha (x))\Lambda^{~n}_{s}(\alpha (x))-d\alpha^{mn}(x) .
\label{oneformvari}
\eea
For infinitesimal transformations, the local Lorentz transformations are $\Lambda^{~m}_{n}(\alpha (x)) = \delta^{~m}_{n} + \alpha^{~m}_{n}(x)$ and the spinor transformations are $D_{\alpha}^{(\frac{1}{2},0)\beta} (\alpha (x))= \delta_\alpha^{~\beta} +\frac{i}{4} \alpha_{mn} (x)\left(\sigma^{mn} \right)_\alpha^{~~\beta}$ and $D_{~~~~~~~\dot\beta}^{(0,\frac{1}{2})\dot\alpha} (\alpha (x))= \delta^{\dot\alpha}_{~\dot\beta} + 
\frac{i}{4} \alpha_{mn}\left(\bar\sigma^{mn} \right)^{\dot\alpha}_{~~\dot\beta}$
, while the infinitesimal local ${\cal SP}_4$ transformations of the gravitational one-forms take the form
\bea
\hat{E}^{\prime m} &=& \hat{E}^m + \hat{\gamma}^{mn}\epsilon_n +2i\left( \xi \sigma^m \bar{\hat\psi} - \hat\psi \sigma^m \bar\xi \right) -\alpha^{mn}\hat{E}_n -d\epsilon^m \cr
\hat\psi^{\prime\alpha} &=& \hat\psi^\alpha -\frac{i}{4}\alpha_{mn}(\hat\psi\sigma^{mn})^\alpha+\frac{i}{4}\hat\gamma_{mn}(\xi\sigma^{mn})^\alpha -d\xi^\alpha \cr
\bar{\hat\psi}_{\dot\alpha}^\prime &=& \bar{\hat\psi}_{\dot\alpha} -\frac{i}{4}\alpha_{mn}(\bar{\hat\psi} \bar\sigma^{mn})_{\dot\alpha} +\frac{i}{4}\hat\gamma_{mn}(\bar{\xi} \bar\sigma^{mn})_{\dot\alpha} -d\bar\xi_{\dot\alpha} \cr
\hat\gamma^{\prime mn} &=& \hat\gamma^{mn} + ( \alpha^{mr}\hat\gamma^n_{~~r} - \alpha^{nr} \hat\gamma^m_{~~r}) -d\alpha^{mn}.
\eea

Using the Feynman formula for the variation of an exponential operator in conjunction with the Baker-Campbell-Hausdorff formulae, the individual one-forms appearing in the above decomposition of the covariant Maurer-Cartan one-form are secured as 
\bea
\omega^m &=& dx^m -i[\lambda \sigma^m (d\bar\lambda +2\bar\psi) - (d\lambda +2\psi) \sigma^m \bar\lambda] +E^m +\frac{1}{4}\gamma_{rs}\lambda(\sigma^m \bar\sigma^{rs} +\sigma^{rs} \sigma^m)\bar\lambda \cr
\omega_Q^\alpha &=& d\lambda^\alpha + \psi^\alpha -\frac{i}{4}\gamma_{mn} (\lambda \sigma^{mn})^\alpha \cr
\omega_{\bar{Q}\dot\alpha} &=& d\bar\lambda_{\dot\alpha} + \bar\psi_{\dot\alpha} -\frac{i}{4}\gamma_{mn} (\bar\lambda \bar\sigma^{mn})_{\dot\alpha} \cr
\omega_M^{mn} &=& \gamma^{mn} .
\label{MCOne-form}
\eea

The covariant coordinate differential $\omega^m$ is related to the space-time coordinate differential $dx^\mu$ by the vierbein $e_\mu^{~m}$ so that $\omega^m = dx^\mu e_\mu^{~m}$
\be
e_\mu^{~m}= A_\mu^{~m} +E_\mu^{~m}-2i(\lambda \sigma^m \bar\psi_\mu -\psi_\mu \sigma^m \bar\lambda) +\frac{1}{4}\gamma_\mu^{rs} \lambda(\sigma^m \bar\sigma_{rs} +\sigma_{rs} \sigma^m ) \bar\lambda .
\ee

The one-forms and their covariant derivatives are the building blocks of the locally ${\cal SP}_4$ invariant action.  Indeed a $m^{\rm th}$-rank contravariant local Lorentz and $n^{\rm th}$-rank covariant Einstein tensor, $T^{m_1\cdots m_m}_{\mu_1\cdots \mu_n}$, is defined to transform as \cite{Utiyama:1956sy}
\be
T^{\prime m_1^\prime\cdots m_m^\prime}_{\mu_1^\prime\cdots \mu_n^\prime}(x^\prime) = G_{\mu_1^\prime}^{-1\mu_1}(x)\cdots G_{\mu_n^\prime}^{-1\mu_n}(x) T^{m_1\cdots m_m}_{\mu_1\cdots \mu_n}(x)
\Lambda^{~m_1^\prime}_{m_1} (\alpha (x))\cdots \Lambda^{~m_m^\prime}_{m_m} (\alpha (x)), 
\ee
while a local Lorentz $(m,n)$ rank spinor transforms as
\bea
\Psi^{\prime~~\dot\alpha_1\cdots \dot\alpha_n}_{\alpha_1 \cdots \alpha_m} (x^\prime) &=& D_{\alpha_1}^{(\frac{1}{2},0)\beta_1}(\alpha(x))\cdots D_{\alpha_m}^{(\frac{1}{2},0)\beta_n}(\alpha(x)) \Psi^{~~\dot\beta_1\cdots \dot\beta_n}_{\beta_1 \cdots \beta_m} (x)\cr
 & &\qquad\qquad\qquad D^{(0,\frac{1}{2})\dot\alpha_1}_{\dot\beta_1} (\alpha(x))\cdots D^{(0,\frac{1}{2})\dot\alpha_n}_{\dot\beta_n} (\alpha(x)).
\eea
A mixed tensor-spinor is defined to transform analogously using the above transformation properties of pure quantities.  For example, the vierbein transforms as $e_\mu^{\prime m} (x^\prime) = G_{\mu}^{-1\nu}(x) e_\nu^{~n}(x)\Lambda^{~m}_{n} (\alpha (x))$ while the covariant derivative of the Goldstino transforms as $\omega_{Q\mu\alpha}^\prime (x^\prime) = G_{\mu}^{-1\nu}(x)D_{\alpha}^{(\frac{1}{2},0)\beta}(\alpha(x))\omega_{Q\nu\beta} (x)$.  Hence, the vierbein and its inverse can be used to convert local Lorentz indices into space-time, that is, world indices and vice versa.  Since the Minkowski metric, $\eta_{mn}$, is invariant under local Lorentz transformations the metric tensor $g_{\mu\nu}$
\be
g_{\mu\nu} = e_\mu^{~m} \eta_{mn} e_\nu^{~n} ,
\ee
is a rank 2 Einstein tensor.  It can be used to define covariant Einstein tensors given contravariant ones.  Likewise, the Minkowski metric can be used to define covariant local Lorentz tensors given contravariant ones, while the antisymmetric 2-index symbol,
$\epsilon_{\alpha\beta}$ and $\epsilon^{\alpha\beta}$ and analogously for the dotted indices, can be used to raise, lower and contract spinor indices in the usual fashion.  

Since the  $x^\mu \rightarrow x^{\prime \mu}$ transformation produces the volume element transformation
\bea
d^4 x^\prime &=& d^4 x ~\det{{G}},  
\eea
while $\det{\Lambda} = 1$, it follows that $d^4 x^\prime ~\det{e^\prime} (x^\prime) = d^4 x ~\det{e} (x)$.
Thus an ${\cal SP}_4$ invariant action can be  constructed as
\be
\Gamma = \int d^4 x \det{e(x)} {\cal L}(x),
\ee
where ${\cal L}^\prime (x^\prime) = {\cal L}(x)$ is any invariant Lagrangian.  The invariants that make up such a Lagrangian can be found by contracting tensor indices with the appropriate vierbein, its inverse and the Minkowski metric and spinor indices with the appropriate epsilon symbols.  For example $\omega_{Q\mu}^\alpha g^{\mu\nu}\epsilon_{\alpha\beta} \omega_{Q\nu}^\beta$ is an invariant term which can be used in the construction of the action.

Besides products of the covariant Maurer-Cartan one-forms, their covariant derivatives can also be used to construct invariant terms of the Lagrangian.  The covariant derivative of a general tensor can be defined using the affine and related spin connections.  Consider the covariant derivative of the Lorentz tensor $T^{mn}$
\be
\nabla_\rho T^{mn} = \partial_\rho T^{mn} -\omega_{M\rho r}^{m} T^{rn}-\omega_{M\rho r}^{n} T^{mr} .
\label{covderivtensor}
\ee
Since the spin connection transforms inhomogeneously according to equation (\ref{oneformvari}), the covariant derivative of $T^{mn}$ transforms homogeneously again 
\be
(\nabla_\rho T^{mn})^\prime =G_\rho^{-1\sigma} (\nabla_\sigma T^{rs})\Lambda_r^{~m}\Lambda_s^{~n} .
\ee
Converting the Lorentz index $n$ to a space-time index $\nu$ using the vierbein, the covariant derivative for mixed tensors is obtained
\bea
\nabla_\rho T^{m\nu} &\equiv & e_n^{-1\nu} \nabla_\rho T^{mn} = \partial_\rho T^{m\nu} -\omega_{M\rho}^{mr}T_r^{~\nu} + \Gamma_{\sigma\rho}^\nu T^{m\sigma} ,
\eea
where the spin connection $\omega_{M\rho}^{mn}$ and $\Gamma_{\sigma\rho}^\nu$ are related according to \cite{Utiyama:1956sy}
\be
\Gamma_{\sigma\rho}^\nu = e_n^{-1\nu} \partial_\rho e_\sigma^{~n} -e_n^{-1\nu} \omega_{M\rho}^{nr} e_\sigma^{~s} \eta_{rs}.
\ee
(Note that this relation as well follows from the requirement that the covariant derivative of the vielbein vanishes, $\nabla_\rho e_\mu^{~m} =0$.)
Applying the above to the Minkowski metric Lorentz 2-tensor yields the formula relating the affine connection $\Gamma^\rho_{\mu\nu}$ to derivatives of the metric
\bea
\nabla_\rho \eta^{mn} &=& \partial_\rho \eta^{mn} -\omega_{M\rho r}^{m} \eta^{rn}-\omega_{M\rho r}^{n} \eta^{mr} \cr
 &=& -\omega_{M\rho}^{mn} -\omega_{M\rho}^{nm} =0\cr
 &=& e_\mu^{~m}e_\nu^{~n} \nabla_\rho g^{\mu\nu} \cr
 &=& e_\mu^{~m}e_\nu^{~n}\left( \partial_\rho g^{\mu\nu} +\Gamma_{\sigma\rho}^\mu g^{\sigma\nu} + \Gamma_{\sigma\rho}^\nu g^{\mu\sigma} \right).
\eea
The solution to this equation (for the case that the space is torsionless, the connection is symmetric $\Gamma^\rho_{\mu\nu} = \Gamma^\rho_{\nu\mu}$) yields the affine connection in terms of the derivative of the metric \cite{Utiyama:1956sy} 
\be
\Gamma^\rho_{\mu\nu} = \frac{1}{2}g^{\rho\sigma}\left[ \partial_\mu g_{\sigma\nu} +\partial_\nu g_{\mu\sigma} - \partial_\sigma g_{\mu\nu}\right].
\ee

Finally a covariant field strength curvature two-form can be constructed out of the inhomogeneously transforming spin connection $\omega_{M\mu}^{mn}$
\bea
R^{mn} &=& d\omega_M^{mn} +\eta_{rs} \omega_M^{mr}\wedge \omega_M^{ns} .
\eea
Expanding the forms yields the field strength tensor
\be
R^{mn}_{\mu\nu} = \partial_\mu \omega_{M\nu}^{mn} -\partial_\nu \omega_{M\mu}^{mn}+\eta_{rs} \omega_{M\mu}^{mr} \omega_{M\nu}^{ns} -\eta_{rs} \omega_{M\nu}^{mr} \omega_{M\mu}^{ns} .
\ee
It can be shown that $R^{mn}_{\mu\nu}=e^{-1n\sigma}e_\rho^{~m}R^\rho_{~\sigma\mu\nu}$ where $R^\rho_{~\sigma\mu\nu}$ is the Riemann curvature tensor
\be
R^\rho_{~\sigma\mu\nu} = \partial_\nu \Gamma^\rho_{\sigma\mu} -\partial_\mu \Gamma^\rho_{\sigma\nu} +\Gamma^\lambda_{\sigma\mu}\Gamma^\rho_{\lambda\nu} -\Gamma^\lambda_{\sigma\nu}\Gamma^\rho_{\lambda\mu} .
\ee
The Ricci tensor is given by $R_{\mu\nu} = R^\rho_{\mu\nu\rho}$ and hence the scalar curvature is an invariant
\be
R = g^{\mu\nu} R_{\mu\nu} = - e^{-1\mu}_m e_n^{-1\nu} R^{mn}_{\mu\nu} .
\ee
In similar fashion the covariant derivatives of spinor one-forms $\Psi_{\mu\alpha}$ and $\bar\Psi_\mu^{~\dot\alpha}$, for example, are defined as
\bea
\nabla_\rho \Psi_{\mu\alpha} &=& \partial_\rho \Psi_{\mu\alpha} +\frac{i}{4}\omega_{M\rho}^{mn} (\sigma_{mn})_\alpha^{~\beta} \Psi_{\mu\beta} -\Gamma^\nu_{\rho\mu}\Psi_{\nu\alpha} \cr
\nabla_\rho \bar\Psi_{\mu}^{~\dot\alpha} &=& \partial_\rho \bar\Psi_\mu^{~\dot\alpha} +\frac{i}{4}\omega_{M\rho}^{mn} (\bar\sigma_{mn})^{\dot\alpha}_{~\dot\beta} \bar\Psi_\mu^{~\dot\beta} -\Gamma^\nu_{\rho\mu}\bar\Psi_\nu^{~\dot\alpha} .
\label{covderivspinor}
\eea

\newsection{The Invariant Action}

The covariant derivatives of the Maurer-Cartan one-forms provide additional building blocks out of which the invariant action is to be constructed.  For example the covariant derivatives of $\omega_{Q\alpha}=dx^\nu \omega_{Q\nu\alpha}$ and $\bar\omega_{\bar{Q}}^{~\dot\alpha}= dx^\nu \bar\omega_{\bar{Q}\nu}^{~~\dot\alpha}$ yield the mixed tensors
\bea
\nabla_\mu \omega_{Q\nu\alpha} &=& \partial_\mu \omega_{Q\nu\alpha} +\frac{i}{4}\gamma_\mu^{mn} (\sigma_{mn})_\alpha^{~\beta} \omega_{Q\nu\beta} -\Gamma^\rho_{\mu\nu}\omega_{Q\rho\alpha} \cr
\nabla_\mu \bar\omega_{\bar{Q}\nu}^{~~\dot\alpha} &=& \partial_\mu \bar\omega_{\bar{Q}\nu}^{~~\dot\alpha} +\frac{i}{4}\gamma_\mu^{mn} (\bar\sigma_{mn})^{\dot\alpha}_{~\dot\beta} \bar\omega_{\bar{Q}\nu}^{~~\dot\beta} -\Gamma^\rho_{\mu\nu}\bar\omega_{\bar{Q}\rho}^{~~\dot\alpha} .
\label{MCcovderiv}
\eea
Thus the invariant action describing spontaneously broken supergravity has the general low energy form
\bea
\Gamma &=& \int d^4 x \det{e} \left\{ \Lambda + \frac{1}{2\kappa^2} R -\frac{i}{2}m_{3/2} \left[\omega_{Q\mu}^{~~~\alpha}\sigma^{\mu\nu\beta}_\alpha\omega_{Q\nu\beta}+\bar\omega_{\bar{Q}\mu\dot\alpha}\bar\sigma^{\mu\nu\dot\alpha}_{~~~~~\dot\beta}\bar\omega_{\bar{Q}\nu}^{~~~\dot\beta} \right]\right.\cr
 & &\left. -\frac{1}{2}M \left.\left[\omega_{Q\mu}^{~~~\alpha}g^{\mu\nu}\omega_{Q\nu\alpha}+\bar\omega_{\bar{Q}\mu\dot\alpha}g^{\mu\nu}\bar\omega_{\bar{Q}\nu}^{~~~\dot\alpha} \right]\right. + Z\epsilon^{\mu\nu\rho\sigma}\omega_{Q\mu} (\sigma^s e_{s\sigma}^{-1}) \nabla_\rho \bar\omega_{\bar{Q}\nu}\right.\cr
 & &\left.+ \omega_{Q\mu} \left[ iZ_1 g^{\mu\nu}\sigma^\rho + iZ_2 g^{\mu\rho} \sigma^\nu +iZ_3 g^{\nu\rho}\sigma^\mu \right]\nabla_\rho \bar\omega_{\bar{Q}\nu}\right\} .
\eea
Higher dimension invariant terms are also possible but are suppressed by additional powers of the largest scale, typically $M_{\rm pl}\sim 1/\kappa$.  The minimal model with the parameters $M$, $Z_1$, $Z_2$ and $Z_3$ set to zero can be consistently quantized~\cite{Rarita:1941mf}.  On the other hand, due to the Higgs mechanism, the parameter $m_{3/2}$ cannot be zero and is an independent scale in the theory.  Hence, the cosmological constant, $\Lambda$, the gravitino mass scale, $m_{3/2}$, which is related to the scale of SUSY breaking, $M_S \sim m_{3/2}$, and the gravitational scale, $M_{\rm pl}$, are all potentially independent scales \cite{Deser:1977uq}. As such, a  minimal nonlinearly realized supergravity action is given by
\bea
\Gamma &=& \int d^4 x \det{e} \left\{ \Lambda + \frac{1}{2\kappa^2} R + Z\epsilon^{\mu\nu\rho\sigma}\omega_{Q\mu} (\sigma^s e_{s\sigma}^{-1}) \nabla_\rho \bar\omega_{\bar{Q}\nu}\right.\cr
 & &\qquad\qquad\qquad\left.-\frac{i}{2}m_{3/2} \left[\omega_{Q\mu}^{~~~\alpha}\sigma^{\mu\nu\beta}_\alpha\omega_{Q\nu\beta}+\bar\omega_{\bar{Q}\mu\dot\alpha}\bar\sigma^{\mu\nu\dot\alpha}_{~~~~~\dot\beta}\bar\omega_{\bar{Q}\nu}^{~~~\dot\beta} \right]\right\} .
\label{effaction}
\eea

Since $\lambda_\alpha$ and $\bar\lambda_{\dot\alpha}$ transform inhomogeneously under the broken local SUSY transformations, we can now fix the unitary gauge defined by  $\lambda_\alpha =0= \bar\lambda_{\dot\alpha}$. So doing, the covariant one-forms take a simplified form\hfill\pagebreak
\bea
\omega^m &=& dx^m +E^m =dx^\mu e_\mu^{~m}  \cr
\omega_Q^{~\alpha} &=& \psi^\alpha =dx^\mu \psi_\mu^{~\alpha}  \cr
\omega_{\bar{Q}\dot\alpha} &=&  \bar\psi_{\dot\alpha}=dx^\mu \bar\psi_{\mu\dot\alpha} \cr
\omega_M^{mn} &=& \gamma^{mn} .
\label{MCOne-formUnitary}
\eea
Note that the $\det{e} $ gives no contribution to the gravitino mass even though it is the source of Goldstino kinetic term in the model with spontaneously broken global SUSY. Instead, the mass of the gravitino, $m_{3/2}$, is a completely new scale arising from an independent monomial, a monomial that vanished in the global case \cite{Deser:1977uq}. This is reminsicent of what transpires when gauging the spontaneously broken isometries of  $AdS_5$ space on an embedded $AdS_4$ manifold \cite{Clark:2005ec}. In that case, the spectrum contains a massive Abelian vector whose mass is an independent scale. On the other hand, this realization of the Higgs mechanism is strikingly different from what occurs when gauging internal symmetries. In that case, when the symmetry is made local, the Nambu-Goldstone boson kinetic term gets replaced by the square of the covariant derivative containing the vector connection.  In unitary gauge, the Nambu-Goldstone field vanishes leaving the residual vector mass term whose scale is set by the Nambu-Goldstone decay constant, a scale already present in the global model.  Here the former Goldstino kinetic energy term becomes a cosmological constant term.

Thus, in unitary gauge, the action, equation (\ref{effaction}), reduces to that of a massive gravitino field coupled to a gravitational field with cosmological constant
\be
\Gamma = \int d^4 x \det{e} \left\{ \Lambda + \frac{1}{2\kappa^2} R + Z\epsilon^{\mu\nu\rho\sigma}\psi_{\mu} (\sigma^s e_{s\sigma}^{-1}) \nabla_\rho \bar\psi_{\nu}
-\frac{i}{2}m_{3/2} \left[\psi_{\mu} \sigma^{\mu\nu}\psi_{\nu}+\bar\psi_{\mu}\bar\sigma^{\mu\nu}\bar\psi_{\nu}\right]\right\} .
\ee

\newsection{Invariant Coupling To Matter}

As discussed in section 2, matter fields can be characterized by their
Lorentz group (with generators $M^{mn}$) transformation properties.  
Each matter field, $M(x)$, transforms under $G$ as
\be
M^\prime (x^\prime ) \equiv \tilde{h} M(x) ,
\ee
where $\tilde{h}$ is given by
\be
\tilde{h} = e^{\frac{i}{2}\alpha_{mn}(x) \tilde{M}^{mn}},
\ee
with $\tilde{M}^{mn}$ the matrix for the corresponding matter field representation of the Lorentz algebra. For example, a scalar
field, $S(x)$, is in the trivial representation of the Lorentz group, 
$\tilde{M}^{mn}=0$, while fermion fields, $\psi_\alpha (x)$ or $\bar{\psi}_{\dot\alpha}$, carry the (1/2, 0) spinor representation, $(\tilde{M}^{mn})_{\alpha}^{~\beta}=1/2 (\sigma^{mn})_{\alpha}^{~\beta}$, or the (0, 1/2) spinor representation, $(\tilde{M}^{mn})^{\dot\alpha}_{~\dot\beta}=1/2 (\bar\sigma^{mn})^{\dot\alpha}_{~\dot\beta}$ . The covariant derivative for the matter field is defined using the Maurer-Cartan spin connection one-form (c.f. Eqs. (\ref{covderivtensor}) and (\ref{covderivspinor})) as \cite{Clark:2002bh}
\be
\nabla M \equiv ( d +\frac{i}{2}\omega_{M}^{mn} \tilde{M}_{mn} ) M 
\ee
so that it has the same transformation properties as the matter field itself,
\be
(\nabla M)^\prime (x^\prime) = \tilde{h} \nabla M(x) .
\label{Mprime}
\ee
Expanding the covariant derivative one-form in terms of space-time coordinate differentials, $dx^\mu$, the component form of the covariant derivative is given by
\be
\nabla_\mu M = \left( \partial_\mu +\frac{i}{2} \gamma_\mu^{~mn} \tilde{M}_{mn} \right) M 
\ee
and exhibits the ${\cal SP}_4$ transformation law
\be
\left( \nabla_\mu M\right)^\prime (x^\prime) = \tilde{h} G_\mu^{-1 \nu}
\nabla_\nu M
(x) .
\ee

The definition of the covariant derivative can be extended when the
matter fields also carry a representation of a local internal symmetry group
${\cal G}$, \cite{Clark:2002bh}, \cite{Clark:2002rp}, so that
\be
M^{\prime a}(x) = (U(\epsilon))^a_{~b} M^b (x) ,
\ee
where the representation matrix
\be
(U(\epsilon))^a_{~b} = (e^{ig\epsilon^A (x) T^A})^a_{~b},
\ee
is given in terms of the local transformation parameters
$\epsilon^A (x)$ and the gauge coupling constant $g$.
The generator representation 
 matrices $(T^A)^a_{~b}$, $A= 1,2,\ldots, {\rm dim}{\left[\cal 
G\right]}$ satisfy the associated Lie algebra $[T^A , T^B ] = if^{ABC} T^C$,
and are normalized so that ${\rm Tr}{[T^A T^B]} = 1/2 \delta^{AB}$.
In order to extend the invariance of the action to include gauge 
transformations,
the Yang-Mills gauge potential one-form, 
\be
A(x) = dx^\mu A_\mu (x) = dx^\mu (iT^A A^A_\mu (x)) .
\ee
must be
introduced. Under ${\cal SP}_4$-transformations, this one-form is invariant: $A^\prime
(x^\prime) = A(x)$, and thus the gauge field transforms as a
coordinate differential
\be
A_\mu^\prime (x^\prime) = G_\mu^{-1 \nu} A_\nu (x) ,
\ee
while under ${\cal G}$-transformations the Yang-Mills field transforms as a
gauge connection
\be
A^\prime = U(\epsilon)AU^{-1}(\epsilon) +\frac{1}{g}
(dU(\epsilon))U^{-1}(\epsilon) .
\ee
Thus the gauge and super-Poincar\'e covariant derivative of the matter
field is secured as
\be
\nabla M = [d +\frac{i}{2}\omega^{mn}_M \tilde{M}_{mn} -gA ] M
\ee
so that, under super-Poincar\'e transformations the covariant derivative transforms
identically to $M$, $(\nabla M)^\prime (x^\prime) = \tilde{h} (\nabla M)(x)$, while under gauge transformations the covariant derivative carries 
the same matter field representation of ${\cal G}$ as $M$:
\be
(\nabla M)^\prime = U(\epsilon) (\nabla M) .
\ee
The matter field covariant derivative can be expanded in terms of the space-time
coordinate differentials $dx^\mu$ giving 
\be
\nabla_\mu M = \left( \partial_\mu +\frac{i}{2}\gamma_{\mu}^{~mn} \tilde{M}_{mn} -gA_\mu\right)M .
\ee
The fully covariant derivatives for the scalar, $S(x)$, and fermion,
$\psi_\alpha(x)$ and $\bar\psi^{\dot\alpha} (x)$, matter fields have the explicit form
\bea
(\nabla_\mu S)^a &=& \partial_\mu S^a -igA_\mu^A (T^A)^a_{~b} 
S^b \cr
(\nabla_\mu \psi)^a_\alpha &=& \partial_\mu \psi_\alpha^a +\frac{i}{4}
\gamma_\mu^{~mn}
(\sigma_{mn})_{\alpha}^{~\beta} \psi_\beta^a -ig A_\mu^A
(T^A)^a_{~b} \psi^b_\alpha \cr
(\nabla_\mu \bar\psi)^{\dot\alpha a} &=& \partial_\mu \bar\psi^{\dot\alpha a} +\frac{i}{4}\gamma_\mu^{~mn}
(\bar\sigma_{mn})^{\dot\alpha}_{~\dot\beta} \bar\psi^{\dot\beta a} -ig A_\mu^A
(T^A)^a_{~b} \bar\psi^{\dot\alpha b} .
\eea

The Yang-Mills field strength two-form, $F$, is defined as $F\equiv dA + g 
A\wedge A$ .
As a two-form, $F$ is invariant under ${\cal SP}_4$-transformations while under 
${\cal G}$-transformations it is in the adjoint representation
$F^\prime = U(\epsilon) F U^{-1}(\epsilon)$ .
Expanding $F$ in terms of the coordinate differential basis $dx^\mu$,
$F=\frac{1}{2} dx^\nu \wedge dx^\mu (iT^A F^A_{\mu\nu})$, the space-time index field strength tensor is obtained as
\be
F^A_{\mu\nu} = \partial_\mu A_\nu^A -\partial_\nu A^A_\mu + g f^{ABC} A_\mu^B A_\nu^C .
\ee

 A generic nonlinearly realized supergravity and gauge invariant matter field action can be constructed as
\be
\Gamma_{\rm matter} = \int d^4 x \det{e}~{\cal L}_{\rm matter},
\ee
where the fully invariant matter field Lagrangian ${\cal L}_{\rm matter}$ takes the form
\be
{\cal L}_{\rm matter}= {\cal L}_{\rm matter}(M, \nabla_\mu M, \omega_Q, \bar\omega_{\bar{Q}}, \nabla_\mu \omega_Q, \nabla_\mu\bar\omega_{\bar{Q}}, e_\mu^{~m}, R_{\mu\nu\rho\sigma}, F_{\mu\nu}^A),
\ee
where ${\cal L}_{\rm matter}$ is any SUSY and gauge invariant function of the basic building blocks which consist of the vierbein, $e_\mu^{~m}$, the  fermionic Maurer-Cartan one forms, $\omega_{Q\mu}^\alpha$ and $\bar\omega_{\bar{Q}\mu\dot\alpha}$, their covariant derivatives as given in equation (\ref{MCcovderiv}), the Riemann tensor, $R_{\mu\nu\rho\sigma}$, the matter fields, $M$, their covariant derivatives, $\nabla_\mu M$, the gauge field strength tensor, $F_{\mu\nu}^A$, and higher covariant derivatives of all quantities.  Combined with the pure supergravity action of equation (\ref{effaction}), the low energy effective action describing the dynamics of the light matter and gauge fields along with the graviton and massive Goldstino/gravitino fields is given by
\be
\Gamma_{\rm eff} = \int d^4 x \det{e}~{\cal L}_{\rm eff},
\ee
where again the fully SUSY and gauge invariant effective Lagrangian has the generic form
\be
{\cal L}_{\rm eff}= {\cal L}_{\rm eff}(M, \nabla_\mu M, \omega_Q, \bar\omega_{\bar{Q}}, \nabla_\mu \omega_Q, \nabla_\mu\bar\omega_{\bar{Q}}, e_\mu^{~m}, R_{\mu\nu\rho\sigma}, F_{\mu\nu}^A).
\ee

It proves convenient to catalog the terms in the effective Lagranian, 
${\cal L}_{\rm eff}$, by an expansion in the number of Goldstino/gravitino fields which appear after the Goldstino and gravitino fields are set to zero in the fully covariant derivatives and in the vierbein \cite{Clark:1997aa}.  This is tantamount to counting the number of factors of the fermionic Mauer-Cartan one-forms in each expression.  So doing, the effective action has the expansion
\be
{\cal L}_{\rm eff}= \left[{\cal L}_{(0)}+ {\cal L}_{(1)}+ 
{\cal L}_{(2)}+\cdots \right] ~,
\ee
where the subscript $n$ on ${\cal L}_{(n)}$ denotes that each independent 
invariant operator in that set begins with $n$ factors of $\omega_{Q\mu}^\alpha$ and $\bar\omega_{\bar{Q}\mu\dot\alpha}$, or equivalently, with $n$ Goldstino/gravitino fields.  

${\cal L}_{(0)}$  consists of all gauge and SUSY invariant operators 
made only from the vierbein and the light matter and gauge fields and their SUSY covariant derivatives. Thus any Goldstino/gravitino field 
appearing in ${\cal L}_{(0)}$ arises only from 
higher dimension terms in the matter covariant derivatives and/or the field 
strength tensor and vierbein.  For instance, taking the gravitino to be the lightest supersymmetric partner, then ${\cal L}_{(0)}$ has the form
\be
{\cal L}_{(0)} = {\cal L}_{\rm matter} + {\cal L}_{\rm Y-M}, 
\ee
where the matter field action is given by
\bea
{\cal L}_{\rm matter} &=& {\rm Tr}\left[ (\nabla_\mu S)^\dagger g^{\mu\nu}
(\nabla_\nu
S)\right] - V(S) \cr
 & &\qquad +i\psi \sigma^m e_m^{-1 \mu} \nabla_\mu \bar\psi - \psi m \psi -\bar\psi \bar{m} \bar\psi + Y(S,\psi \psi, \bar\psi \bar\psi), 
\label{matlag}
\eea
so that ${\cal L}_{\rm matter}$ includes any possible globally ${\cal G}$-invariant scalar field potential $V(S)$,  
fermion mass terms $\psi m \psi$ and $\bar\psi \bar{m} \bar\psi$, and generalized Yukawa couplings $Y(S,\psi\psi , \bar\psi \bar\psi)$.  The fully invariant Yang-Mills Lagarangian ${\cal L}_{\rm Y-M}$ is 
\be
{\cal L}_{\rm Y-M} = -\frac{1}{2} {\rm Tr}{[F_{\mu\nu} g^{\mu\rho} g^{\nu\sigma} F_{\rho\sigma}]}.
\label{ymlag}
\ee
Note that the coefficients of these terms are fixed by the normalization of the 
gauge and matter fields, their masses and self-couplings. That is, by the normalization of the Goldstino/gravitino independent Lagrangian.  For the case where the gravitino is the only supersymmetric partner whose mass is below the electroweak scale, the matter and gauge field terms are just those of the Standard Model and the normalization of these terms is just that given by the Standard Model.  

The ${\cal L}_{(1)}$ terms in the effective Lagrangian begin with the 
couplings of the non-Goldstino/gravitino fields to only a single Goldstino/gravitino.  The general form of these terms is given by
\be
{\cal L}_{(1)}= \frac{1}{\sqrt{F}}[\omega_{Q\mu}^\alpha Q_{\alpha}^\mu 
+ \bar 
Q_{\dot\alpha}^\mu \bar\omega_{\bar{Q}\mu}^{\dot\alpha}] ,
\ee
where $ Q_{\alpha}^\mu$ and $\bar Q_{\dot\alpha}^\mu$ 
contain the light field contributions to the conserved gauge invariant supersymmetry currents.  That is, it is the term in the effective Lagrangian which involves linear coupling to the Goldstino/gravitino fields.  The Lagrangian ${\cal L}_{(1)}$ describes processes involving the emission or absorption of a single gravitino.  If the gravitino is the lightest supersymmetric partner, the lowest mass dimension $SU(3)\times SU(2) \times U(1)$ invariant terms contributing to ${\cal L}_{(1)}$ are the gravitino-lepton number and R parity conserving terms given by
\be
{\cal L}_{(1)} = \sum_f \frac{c_f}{M_S} \omega_{Q\mu}^\alpha (\sigma^{\mu\nu})_\alpha^{~\beta}[ l_{f\beta}^a (\nabla_\nu \phi)^b \epsilon_{ab}] + h.c.  ,
\label{lepton}
\ee
where $l_{f\alpha} =(\nu_{e_f}~~ e^{-}_f)^{T}_{\alpha}$ is the lepton doublet for family $f$ in the $(1, 2,-\frac{1}{2})$ representation of $SU(3)\times SU(2) \times U(1)$, $\phi = (\phi_+ ~~\phi_0)^T$ is the Higgs doublet in the $(1, 2,+\frac{1}{2})$ representation and $c_f$ are generation dependent effective coupling constants.  In the nonlinearly realized global SUSY case, the processes controlled by such terms involve only the helicity $\pm \frac{1}{2}$ modes of the gravitino.  By means of the equivalence theorem at high energy, they were found from the corresponding Goldstino amplitudes and were delineated and investigated in \cite{Antoniadis:2004se}.  Expanding the Lagrangian, equation (\ref{lepton}), in terms of the component fields and transforming to the unitary gauge for the gravitino and Higgs multiplet so that $\omega_{Q\mu} =\psi_\mu$, $\bar\omega_{\bar{Q}\mu}= \bar\psi_\mu$ and $\phi = (0~~\frac{1}{\sqrt{2}}(v+H))^T$, this now yields the interactions involving all helicities of the gravitino
\bea
{\cal L}_{(1)} &=& -\sum_f \frac{c_f}{M_S}\psi_\mu \sigma^{\mu\nu}\left\{\left[ \frac{1}{\sqrt{2}} \nu_{e_f} \partial_\mu H + h.c. \right] \right.\cr
 & &\left. +\left[ \frac{1}{\sqrt{2}} M_Z \nu_{e_f} Z_\mu + h.c. \right]
 +\left[ i M_W e^-_f W_\mu^+ + h.c.\right] \right.\cr
 & &\left. +\left[ \frac{i}{\sqrt{2}}\frac{e}{\sin{2\theta_W}} \nu_{e_f} Z_\mu H + h.c. \right] 
+\left[i \frac{e}{2\sin{\theta_W}}e^-_f W^+_\mu H + h.c. \right]\right\} .
\eea
The phenomenological consequences of such interaction terms were investigated \cite{Antoniadis:2004se} in the Goldstino/ helicity $\pm\frac{1}{2}$ gravitino case.

Finally the remaining terms in the effective Lagrangian all contain  
two or more Goldstino/gravitino fields.  In 
particular, ${\cal L}_{(2)}$ begins with the coupling of two Goldstino/gravitino 
fields to matter or gauge and gravitational fields.  The lowest dimension such terms, bilinear in $\omega_Q-\bar\omega_{\bar{Q}}$, have the form  
\bea
{\cal L}_{(2)} &=& \frac{1}{F^2}\omega_{Q\mu}^\alpha \bar\omega_{\bar{Q}\nu}^{\dot\alpha} M^{\mu\nu}_{1\alpha\dot\alpha} +\frac{1}{
F^2}
\omega_{Q\mu}^\alpha \stackrel{\leftrightarrow}{\nabla}_\rho \bar\omega_{\bar{Q}\nu}^{\dot\alpha} M^{\mu\nu\rho}_{2\alpha\dot\alpha}\cr
 & &  + \frac{1}{F^2}{\nabla}_\rho
\left[ \omega_{Q\mu}^\alpha \bar\omega_{\bar{Q}\nu}^{\dot\alpha}\right] M^{\mu\nu\rho}_{3\alpha\dot\alpha},
\eea
where the composite operators that contain matter, gauge and gravitational fields are denoted by the $M_i$.  They can be enumerated by their operator dimension, Lorentz 
structure and field content.  Additional discussion of these couplings in the global SUSY case can be found in reference \cite{Clark:1997aa}.

There is another useful one-form basis in which to express the
derivatives and gauge fields \cite{Clark:2002rp}.  The basis consists of the fully covariant coordinate differentials $\omega^m = dx^\mu e_\mu^{~m}$.  
The exterior derivative can be expanded in this basis as 
$d=dx^\mu \partial_\mu = \omega^m {\cal D}_m = \omega^m e_m^{-1\mu} \partial_\mu$, while the gauge field one-form has the analogous expansion
\be
A = dx^\mu A_\mu = \omega^m A_m .
\ee
As previously noted, the covariant basis $\omega^m$ transforms
according to the vector representation of the D=4 (local) Lorentz structure 
group,
$\omega^{\prime~m} = \omega^n \Lambda_n^{~m}$.
So the covariant gauge
field transforms analogously, $A^\prime_m (x^\prime) = \Lambda_m^{-1 n} A_n (x)$,
with the covariant derivative transforming as ${\cal
D}_m^\prime = \Lambda_m^{-1 n} {\cal D}_n$. The invariant interval can be expressed in each of these bases using the metric specific to each as 
\be
ds^2 = dx^\mu g_{\mu\nu} dx^\nu = \omega^m \eta_{mn} \omega^n ,
\ee
with $\eta_{mn}$ the flat tangent space Minkowski metric.

In the new basis, the matter field covariant derivatives take the form
\bea
(\nabla_m S)^a &=& {\cal D}_m S^a -igA_m^A (T^A)^a_{~b} 
S^b \cr
(\nabla_m \psi)^a_\alpha &=& {\cal D}_m \psi_\alpha^a +\frac{i}{4}
\gamma_m^{~rs}
(\sigma_{rs})_{\alpha}^{~\beta} \psi_\beta^a -ig A_m^A
(T^A)^a_{~b} \psi^b_\alpha \cr
(\nabla_m \bar\psi)^{\dot\alpha a} &=& {\cal D}_m \bar\psi^{\dot\alpha a} +\frac{i}{4}\gamma_m^{~rs}
(\bar\sigma_{rs})^{\dot\alpha}_{~\dot\beta} \bar\psi^{\dot\beta a} -ig A_m^A
(T^A)^a_{~b} \bar\psi^{\dot\alpha b} ,
\eea
with $\gamma_m^{rs} = e_m^{-1\mu}\gamma_\mu^{~rs}$.  With these
replacements, the fully invariant matter field Lagrangian, equation 
(\ref{matlag}), can be written as
\bea
{\cal L}_{\rm matter} &=& {\rm Tr}\left[ (\nabla_m S)^\dagger \eta^{mn}
(\nabla_n
S)\right] - V(S) \cr
 & &\qquad +i\psi \sigma^m  \nabla_m \bar\psi - \psi m \psi -\bar\psi \bar{m} \bar\psi + Y(S,\psi \psi, \bar\psi \bar\psi).
\label{matlag2}
\eea

The Yang-Mills action can also be recast in this new basis. The Yang-Mills fields  have the modified gauge variation
\be
A^\prime_m = U(\epsilon)A_m U^{-1}(\epsilon) +\frac{1}{g} ({\cal D}_m
U(\epsilon))U^{-1}(\epsilon) 
\ee
while the field strength tensor takes the form
\be
F^A_{mn} = {\cal D}_m A_n^A - {\cal D}_n A^A_m + g f^{ABC} A_m^B A_n^C  .
\ee
Consequently, in this basis, the fully invariant Yang-Mills
Lagrangian, equation (\ref{ymlag}),  becomes
\be
{\cal L}_{\rm Y-M} = -\frac{1}{2} {\rm Tr}{[F_{mn} \eta^{mr} \eta^{ns}
F_{rs}]} .
\ee

~\\
~\\
\noindent The work of TEC and STL was supported in part by the U.S. Department of Energy under grant DE-FG02-91ER40681 (Task B) while MN was supported by the Japan Society for the Promotion of Science under the Post-Doctoral Research Program.  MN would like to thank the theoretical physics group at Purdue University and TtV would like to thank the theoretical physics groups at Purdue University and the Tokyo Institute of Technology for their hospitality during visits while this work was being completed.

\end{document}